\documentclass[AMA,Times1COL]{WileyNJDv5} %STIX1COL,STIX2COL,STIXSMALL
\usepackage[nameinlink]{cleveref} % Has to be loaded after hyperref package
\articletype{Article Type}%
\usepackage{subfig}
\usepackage{threeparttable}

\received{}
\revised{}
\accepted{}
\journal{}
\volume{}
\copyyear{2023}
\startpage{1}

\begin{document}
	
	\title{A simple sensitivity analysis method for unmeasured confounders via linear programming with estimating equation constraints}
	
	\author[1]{Chengyao Tang}
	
	\author[1,2]{Yi Zhou}
	
	\author[3]{Ao Huang}
	\author[1,4]{Satoshi Hattori*}
	
	\authormark{Tang \textsc{et al.}}
	\titlemark{Bounds for the average treatment effect}
	
	\address[1]{\orgdiv{Department of Biomedical Statistics, Graduate School of Medicine}, \orgname{Osaka University}, \orgaddress{\state{Osaka}, \country{Japan}}}
	
	\address[2]{\orgdiv{Beijing International Center for Mathematical Research}, \orgname{Peking University},\orgaddress{\state{Beijing}, \country{China}}}
	
	\address[3]{\orgdiv{Department of Medical Statistics}, \orgname{University Medical Center G{\"o}ttingen}, \orgaddress{\state{G{\"o}ttingen}, \country{Germany}}}
	
	\address[4]{\orgdiv{Integrated Frontier Research for Medical Science Division, Institute for Open and Transdisciplinary Research Initiatives (OTRI)}, \orgname{Osaka University}, 
		\orgaddress{\state{Osaka}, \country{Japan}}}

	\corres{Satoshi Hattori, Department of Biomedical Statistics, Graduate School of Medicine, Osaka University, Osaka, Japan.\\
	\email{hattoris@biostat.med.osaka-u.ac.jp}}

	%\fundingInfo{Text}
	%\JELinfo{ejlje}
	
	\abstract[Abstract]{In estimating the average treatment effect in observational studies, the influence of confounders should be appropriately addressed. To this end, the propensity score is widely used. If the propensity scores are known for all the subjects, bias due to confounders can be adjusted by using the inverse probability weighting (IPW) by the propensity score. Since the propensity score is unknown in general, it is usually estimated by the parametric logistic regression model with unknown parameters estimated by solving the score equation under the strongly ignorable treatment assignment (SITA) assumption. Violation of the SITA assumption and/or misspecification of the propensity score model can cause serious bias in estimating the average treatment effect. To relax the SITA assumption, the IPW estimator based on the outcome-dependent propensity score has been successfully introduced. However, it still depends on the correctly specified parametric model and its identification. In this paper, we propose a simple sensitivity analysis method for unmeasured confounders. In the standard practice, the estimating equation is used to estimate the unknown parameters in the parametric propensity score model. Our idea is to make inference on the average causal effect by removing restrictive parametric model assumptions while still utilizing the estimating equation. Using estimating equations as constraints, which the true propensity scores asymptotically satisfy, we construct the worst-case bounds for the average treatment effect with linear programming. Different from the existing sensitivity analysis methods, we construct the worst-case bounds with minimal assumptions. We illustrate our proposal by simulation studies and a real-world example.}
	

	\maketitle
	
	%\renewcommand\thefootnote{}
       % \footnotetext{\textbf{Abbreviations:}ATE, average treatment effect; IPW, inverse probability weighting; PS, propensity score.}
	
	\renewcommand\thefootnote{\fnsymbol{footnote}}
	\setcounter{footnote}{1}
	
	\section{Introduction}\label{sec1}
	In observational studies, it is always crucial to adjust influence of confounders in estimating the average treatment effect (ATE). If all the confounders are observed and satisfy the strongly ignorable treatment assignment (SITA) assumption,\cite{rubin1978bayesian,rosenbaum1983central} one can adjust the effects of confounders by using the propensity score. With the propensity score, inverse probability weighting (IPW)\cite{hirano2001,austin2015} is a popular approach. The IPW method constructs weights on the observations of each subject, and then the ATE can be identified by comparing the weighted outcomes of two groups.\cite{lunceford2004} In practice, the propensity score is unknown. Then, the estimation of the propensity score usually relies on a parametric model such as the logistic regression under the SITA assumption. In most observational studies, it is untestable and implausible that there are no unmeasured confounders, and then the SITA assumption may fail to hold. Using the outcome-dependent propensity score is an option to make inference without the SITA assumption.\cite{hernan2004structural,petersen2014causal} By incorporating the outcome variable in the model of the propensity score, we can make inference on the ATE without the SITA assumption. In general, the outcome-dependent propensity score is estimated by a parametric logistic regression model with the observed confounders and the outcome as explanatory variables. Thus, model misspecification is still of concern in the estimation of the outcome-dependent propensity score. Moreover, it has an unidentifiability issue;\cite{scharfstein1999adjusting,miao2016identifiability} that is, the estimating equation cannot determine the unknown parameters uniquely in the outcome-dependent propensity score. Then, the outcome-dependent propensity score cannot solve the issue of unmeasured confounders completely.  \par 
 
	Sensitivity analysis is a useful tool to assess the potential impact of unmeasured confounders, and many sensitivity analysis methods have been developed. With  the substantially increasing applications of the propensity score methods in the analysis of observational studies, there is a growing interest in employing sensitivity analysis methods in real-data analyses. Typical sensitivity analysis approaches involve formulating additional assumptions with regards to the relationships among unmeasured confounders, treatment assignments, and outcomes. These assumptions often take the form of plausible values for parameters that cannot be directly estimated from the observed data and must be set by analysts. Rosenbaum and Rubin \cite{rosenbaum1983assessing} and Lin et al. \cite{lin1998assessing} modeled the mechanism of confounding with both the measured and unmeasured confounders and then estimated the treatment effect parameter of interest. Alternatively, Cornfield et al. \cite{cornfield1959smoking} and Ding and Vanderweele \cite{ding2016sensitivity} developed methods to construct the bounds for the treatment effects to quantify the magnitude of the unmeasured confounders. These bounds were designed to elucidate the extent to which unmeasured confounders could influence observed causal estimates. Particularly, when the sensitivity parameters were expressed as risk ratios, the E-value \cite{vanderweele2017sensitivity} was introduced and has become a pivotal quantity in the realm of causal inference in observational studies. While the E-value can provide a bound without any model specification, the estimand is restrictive and the bound is likely to be wide, which can lead to inefficiency in sensitivity analysis. \par 
	
	For the sensitivity analysis approaches based on the IPW method to estimate the ATE, Li et al. \cite{li2011} modeled the mean between-group differences of potential outcomes to correct bias in the presence of unmeasured confounders. Shen et al.\cite{shen2011sensitivity} proposed an IPW-based sensitivity analysis method by using two parameters, the variance of the multiplicative errors in the estimated propensity score and its correlations with the potential outcomes, to quantify the bias due to unmeasured confounders. Lu and Ding \cite{lu2023flexible} extended the method of Li et al.\cite{li2011} into a more flexible sensitivity analysis framework, which can handle the IPW, outcome regression, and doubly robust estimators. In addition, Zhao et al. \cite{zhao2019} constructed bounds for the ATE based on the IPW estimators by incorporating a marginal sensitivity model.\cite{tan2006} Dorn and Guo \cite{dorn2022} further refined this method and gave sharper bounds. These sensitivity analysis methods can address the impacts of violation of the SITA assumption by quantifying potential biases; however, they rely on untestable parametric assumptions on the departure from the SITA assumption, and it is practically difficult to set a relevant magnitude of the departure. \par

	In this paper, a simple sensitivity analysis framework for unmeasured confounders is proposed. In the standard process of the confounder adjustment with the outcome-dependent propensity score, a parametric model for the propensity score is assumed, and an estimating equation is introduced to estimate its unknown parameters. Instead of determining a unique model for the outcome-dependent propensity score, we construct bounds for the ATE by considering possible propensity scores. We realize it by removing the parametric model for the propensity score, but still relying on the estimating equation. We introduce an optimization problem constrained by the estimating equation, which the true propensity score asymptotically satisfies. The worst-case bounds for the ATE can be obtained by solving a linear programming problem. Different from the existing sensitivity analysis methods, the proposed worst-case bounds do not rely on strong assumptions. By increasing the dimension of the estimating equations involving many covariates, one can make the bounds further narrow. Compared with existing sensitivity analysis methods, the proposed method offers the following advantages. First, the proposed method can provide worst-case bounds with minimal assumptions. Second, since the proposed method is free from the estimated propensity score under the SITA assumption, its misspecification does not matter. Finally, our method exhibits computational efficiency as the optimization problem can be solved by linear programming.\par 
	The rest of this paper is organized as follows. In Section \ref{sec:m}, we introduce the basic notations and the standard methods with the parametric propensity score. In Section \ref{sec:sa}, some existing sensitivity analysis methods for the IPW estimator are reviewed. In Section \ref{sec:p}, the proposed method for sensitivity analysis is introduced. We investigate the performance of the proposed method on simulated datasets in Section \ref{sec:s}, and illustrate it on a real-world example in Section \ref{sec:a}. In Section \ref{sec:d}, we provide a concluding discussion to summarize the main findings and contributions of this paper.
	
	\section{Estimation with the parametric propensity score}\label{sec:m}
	\subsection{Notations and the standard propensity score analysis}\label{sec2.1}
	In this paper, we consider to estimate the ATE for the overall mean over the population in an observational study with two treatment groups. Let $Z$ be the treatment assignment: $Z=1$ if the subject  is in the treated (exposed) group and $Z=0$ if in the control group. Let $X$ be a vector of baseline covariates and $Y$ be the observed outcome. We follow Rubin's causal model framework.\cite{rubin1974} Let $Y^{(1)}$ and $Y^{(0)}$ be the potential outcomes if the subjects were assigned to the treated group ($Z=1$) and the control group ($Z=0$), respectively. Suppose the observational study enrolls $n$ subjects, and the observed data $(Y_i, Z_i, X_i)$ for subject $i$ ($i=1,2,\dots,n$) available, which are independent and identically distributed copies of $(Y, Z, X)$. Denote $\mu_1=E[Y^{(1)}]$ and $\mu_0=E[Y^{(0)}]$. The ATE, which is of our primary interest to estimate, is defined by 
	\begin{equation*}
			\psi=\mu_1-\mu_0=E[Y^{(1)}]-E[Y^{(0)}].
	\end{equation*}
	\par 
	In observational studies, owing to the absence of randomization, the potential influence of confounders should be carefully handled in estimating the ATE. The propensity score is widely used to adjust the bias due to confounding. The propensity score is defined by $e(X_i)=p(Z_i=1\mid X_i)$. Various methods, such as stratification, matching, and IPW,\cite{hirano2001,austin2015,robins1994} can be employed to adjust for confounding with the propensity score. The standard propensity analysis is conducted under the following assumptions:
	\begin{assumption}
		Consistency: $Y_i=Z_iY_i^{(1)}+(1-Z_i)Y_i^{(0)}$ \label{as1}.
	\end{assumption}
	\begin{assumption}
		Positivity: there exists a small positive parameter $\delta$ such that $0<\delta\le e(X_i)\le1-\delta$, for all subjects $i$. \label{as2}
	\end{assumption}
	\begin{assumption}
		SITA: $(Y^{(1)}_i,Y^{(0)}_i)\perp\!\!\!\perp Z_i\mid X_i$.\label{as3}
	\end{assumption}
	
	Assumption 3 implies that the bias due to confounding can be adjusted by using $X$ in principle. The SITA assumption is corresponding to the Missing At Random (MAR) in the missing data analysis context. In this paper, we handle situations in which the SITA is violated, which is corresponding to the concept of the Missing Not At Random (MNAR) in the missing data problem. We use the terminologies SITA and MAR exchangeably. In practice, the propensity score is unknown, and then some parametric models such as the logistic regression model is usually assumed. Let $\mbox{logit}(e(X_i;\theta,\alpha))=\theta+\alpha^{\top}X_i$. The unknown parameters are usually estimated by solving the following score equation:
	\begin{equation}
		\sum_{i=1}^n\left(\begin{array}{cc}
			1\\X_i
		\end{array}\right)\left(Z_i-\frac{exp(\theta+\alpha^{\top}X_i)}{1+exp(\theta+\alpha^{\top}X_i)}\right)=0. \label{eq:7}
	\end{equation}
	Let the solution to the score equation for $(\theta,\alpha)$ be denoted by $(\hat{\theta}, \hat{\alpha})$, and $\hat{e}(X_i)=e(X_i; \hat{\theta}, \hat{\alpha})$. Then, we can determine the unique set of propensity scores for all subjects. In this paper, the propensity score estimated under the SITA assumption is called the MAR-based propensity score to avoid confusion; another type of propensity score is introduced in a later section, which is called the outcome-dependent propensity score. The IPW estimator for $\mu_1$ is defined by
	\begin{equation} 
		\hat{\mu}_1=\frac{1}{n}\sum_{i=1}^{n}\frac{Z_iY_i}{\hat{e}(X_i)}. \label{eq:ipw1}
	\end{equation}
	Similarly, we can estimate $\mu_0$ with
	\begin{equation}
		\hat{\mu}_0=\frac{1}{n}\sum_{i=1}^{n}\frac{(1-Z_i)Y_i}{1-\hat{e}(X_i)},\label{eq:ipw0}
	\end{equation}
	and then the ATE $\psi$ is estimated with
	\begin{equation*}
		\hat{\psi}=\hat{\mu}_1-\hat{\mu}_0.
	\end{equation*}\par
	The aforementioned IPW estimator has an unstabilized form, which may suffer from extremely large weights when some propensity scores are very close to one or zero, and then can cause instability in the estimation. The stabilized IPW (SIPW) estimator introduces a stabilization term to the weights, which helps mitigate the impact of extreme weights. Specifically, the SIPW estimator for $\mu_1$ is defined by
	\begin{equation}
		\hat{\mu}_{1,SIPW}=\left(\frac{1}{n}\sum_{i=1}^n\frac{Z_i}{\hat{e}(X_i)}\right)^{-1}\frac{1}{n}\sum_{i=1}^n\frac{Z_iY_i}{\hat{e}(X_i)}. \label{eq:sipw1}
	\end{equation}
	Similarly, we can estimate $\mu_0$ with
	\begin{equation}
		\hat{\mu}_{0,SIPW}=\left(\frac{1}{n}\sum_{i=1}^n\frac{1-Z_i}{1-\hat{e}(X_i)}\right)^{-1}\frac{1}{n}\sum_{i=1}^n\frac{(1-Z_i)Y_i}{1-\hat{e}(X_i)}, \label{eq:sipw0}
	\end{equation}
	 and then the ATE $\psi$ is estimated with
	\begin{equation*}
		\hat{\psi}_{SIPW}=\hat{\mu}_{1,SIPW}-\hat{\mu}_{0,SIPW}. \label{eq:6}
	\end{equation*}
	In this paper, we focus on the SIPW estimator. \par
	If the SITA assumption holds and the model of the propensity score is correctly specified, the ATE is consistently estimated. However, the SITA assumption does not hold in the presence of unmeasured confounders.

	%%where $g(X)$ refers to a pre-specified veter-valued function of covariates. Notably, dimension of $g(X)$ should be no less than that of $\phi$ in order to solve and dertermine unique propensity score for all subjects, so that we can further obtain the IPW estimator $\hat{\mu}_1$.

	\subsection{Estimation with the outcome-dependent propensity score}\label{sec2.2}
	In this section, suppose that the SITA assumption does not necessarily hold in the presence of unmeasured confounder $U$. The estimation of the ATE using the method in Section \ref{sec2.1} is no longer valid.
	%\begin{equation}
	%	\hat{\mu}_1^*=E_n\left\{\frac{ZY}{\hat{s}(X,U)}\right\},\label{eq:8}
	%\end{equation}
	%\begin{equation}
	%	\hat{\mu}_0^*=E_n\left\{\frac{(1-Z)Y}{1-\hat{s}(X,U)}\right\},\label{eq:9}
	%\end{equation}
	%\begin{equation}
	%	\hat{\psi}^*=\hat{\mu}_1^*-\hat{\mu}_0^*.\label{eq:10}
	%\end{equation}
	To address the issue of unmeasured confounders, the outcome-dependent propensity score approach\cite{greenlees1982,andrea2001,verbeke2001} has been successfully introduced.   We define the outcome-dependent propensity scores by $o^1(X_i,Y_i^{(1)})=P(Z_i=1\mid X_i,Y_i^{(1)})$ and $o^0(X_i,Y_i^{(0)})=P(Z_i=1\mid X_i,Y_i^{(0)})$ for subjects in the treated and control group, respectively.
	%\begin{equation}
	%	\mbox{Nonignorable or true propensity score}: s(X,U)=P(Z=1\mid X,U);
	%\end{equation}
	\par
	%However, under MNAR, equation (\ref{eq:7}) is not sufficient to solve such coefficients of parametric model for outcome-dependet propensity score, since for all subjects only one side of treatment assignment is estimable owing to the violation of SITA. Chang and Kott \cite{chang2008,kott2010} proposed to utilize calibration weighting to adjust for nonresponse, in order to estimate coefficients for $Z$. 
	One may consider the logistic regression models for $o^1(X_i,Y_i^{(1)})$ and $o^0(X_i,Y_i^{(0)})$. Let us consider the models $\mbox{logit}{(o^1(X_i,Y_i^{(1)};\theta^1,\alpha^1,\beta^1))}=\theta^1+\alpha^{1\top}X_i+\beta^1 Y_i^{(1)}$ and $\mbox{logit}{(o^0(X_i,Y_i^{(0)};\theta^0,\alpha^0,\beta^0))}=\theta^0+\alpha^{0\top}X_i+\beta^0 Y_i^{(0)}$. The score equation \eqref{eq:7} does not work for estimation of the unknown parameters in these models, since $Y_i^{(z)}$ is observed only for subjects with $Z_i=z$. The unknown parameters in the model of $o^1(X_i,Y_i^{(1)};\theta^1,\alpha^1,\beta^1)$ can be estimated by solving the following estimating equation:
	\begin{equation}
	\sum_{i=1}^ng(X_i)\left(1-\frac{Z_i}{o^1(X_i,Y_i;\theta^1,\alpha^1,\beta^1)}\right)=0,\label{eq:14}
	\end{equation}
	where $g(X)$ is a vector of the same dimensions as $(\theta^1,\alpha^1,\beta^1)$ and the solution to the estimating equation \eqref{eq:14} is denoted by $(\hat{\theta}^1,\hat{\alpha}^1,\hat{\beta}^1)$. Similarly, the unknown parameters in the model of $o^0(X_i,Y_i^{(0)};\theta^0,\alpha^0,\beta^0)$ can be estimated by solving the following estimating equation:
	\begin{equation}
		\sum_{i=1}^ng(X_i)\left(1-\frac{1-Z_i}{1-o^0(X_i,Y_i;\theta^0,\alpha^0,\beta^0)}\right)=0.\label{eq:15}
	\end{equation}
	The dimension of $g(X)$ should be equal to that of $(\theta^0,\alpha^0,\beta^0)$ to obtain a solution. The solution to the estimating equation \eqref{eq:15} is denoted by $(\hat{\theta}^0,\hat{\alpha}^0,\hat{\beta}^0)$. Denote $\hat{o}^1(X_i,Y^{(1)}_i)=o^1(X_i,Y_i;\hat{\theta}^1,\hat{\alpha}^1,\hat{\beta}^1)$ and $\hat{o}^0(X_i,Y^{(0)}_i)=o^0(X_i,Y_i;\hat{\theta}^0,\hat{\alpha}^0,\hat{\beta}^0)$, respectively. We can then estimate $\mu_1$ under the MNAR with 
	\begin{equation*}
		\hat{\mu}^{MNAR}_{1,SIPW}=\left(\frac{1}{n}\sum_{i=1}^n\frac{Z_i}{\hat{o}^1(X_i,Y^{(1)}_i)}\right)^{-1}\frac{1}{n}\sum_{i=1}^n\frac{Z_iY_i}{\hat{o}^1(X_i,Y^{(1)}_i)}.
	\end{equation*}
	Similarly, we can estimate $\mu_0$ under the MNAR with
	\begin{equation*}
		\hat{\mu}^{MNAR}_{0,SIPW}=\left(\frac{1}{n}\sum_{i=1}^n\frac{1-Z_i}{1-\hat{o}^0(X_i,Y^{(0)}_i)}\right)^{-1}\frac{1}{n}\sum_{i=1}^n\frac{(1-Z_i)Y_i}{1-\hat{o}(X_i,Y^{(0)}_i)},
	\end{equation*}
	and then the ATE is estimated with
		\begin{equation*}
		\hat{\psi}^{MNAR}_{SIPW}=\hat{\mu}^{MNAR}_{1,SIPW}-\hat{\mu}^{MNAR}_{0,SIPW}. 
	\end{equation*}

	The SIPW estimator with the outcome-dependent propensity score can consistently estimate the ATE without the SITA assumption as long as the parametric models for the outcome-dependent propensity score are correctly specified. However, estimations with \eqref{eq:14} and \eqref{eq:15} often encounter an unidentifiability issue, wherein the model coefficients obtained through solving the estimating equations may not be uniquely determined. Miao et al.\cite{miao2016identifiability} pointed out that even if the model for the propensity score has a known parametric form, the model is not identifiable without specifying a parametric outcome distribution. A unique solution to the estimating equations is only achieved when both the outcome model and the propensity score model are appropriately specified. Specifically, without additional restrictions or assumptions, sorely solving the estimating equations \eqref{eq:14} is not sufficient to determine the coefficients $(\hat{\theta}^1,\hat{\alpha}^1,\hat{\beta}^1)$ uniquely. Therefore, the outcome-dependent propensity score cannot solve the issue of the unmeasured confounder completely.
	
	%Correspondingly,  
	%	\begin{equation}
	%		\sum_{i=1}^ng(X_i)\left(1-\frac{Z_i}{o^1(X_i,Y^{(1)}_i)}\right)=0 \label{eq:20}
	%	\end{equation} 
	%holds for every outcome-dependet propensity scores $o^1(X_i,Y^{(1)}_i)$. Similarly, in the estimation of $\mu_0$, although sovling the estimating equation (\ref{eq:15}) cannot give unique coeffecients $(\hat{\theta}^0,\hat{\alpha}^0,\hat{\beta}^0)$, 
	%	\begin{equation}
	%		\sum_{i=1}^ng(X_i)\left(1-\frac{1-Z_i}{o^0(X_i,Y^{(0)}_i)}\right)=0 \label{eq:21}
	%	\end{equation} 
	%is asymptotically statisfied. \par

	\section{Existing sensitivity analysis methods}\label{sec:sa}
	%However, the specific values of $o^1(X_i,Y^{(1)}_i)$ and $o^0(X_i,Y^{(0)}_i)$ for all subjects are not necessarily required in the sensitivity analysis. Although the estimating equations (\ref{eq:14}) and (\ref{eq:15}) are not sufficient to determine the unique coefficients of the parametric model for propensity score, under MNAR, it still contains some restrictions and implications on the outcome-dependent propensity score. Moreover, in the sensitivity analysis, our interest is to approach unbiased IPW estimators under MNAR, or to obtain a bound in which the true causal effect can be included, as the impact of the unmeasured confounders, thus assessing the robustness of causal inference in the primary analysis. The exact coefficients of estimated propensity scores are not necessarily required. Indeed, estimations of IPW estimators for ATE usually begin with estimations of propensity scores for all subjects. However, in sensitivity analysis, we contemplate the possibility of bypassing the specific estimation of propensity scores for each individual while still relying on the estimating equations. This motivation can help circumvent the unidentifiability issue and the potential bias resulting from model misspecification. In our proposed sensitivity analysis framework, we will incorporate the estimating equation based on outcome-dependent propensity score without solving it for acquiring its specific model. 
	In this section, we will briefly review some existing sensitivity analysis methods for the IPW estimator.
	
	\subsection{Modeling the mean difference of the potential outcomes}\label{sec2.31}
	Along with the lines of the work by Robins et al.,\cite{robins1999association,robins2000sensitivity} Brumback et al. \cite{brumback2004sensitivity} proposed to quantify the impact of the unmeasured confounders by modeling the mean between-group difference of the potential outcomes, conditional on all observed covariates. The sensitivity function is defined by $c(z,X)=E[Y^{(z)}\mid Z=1,X]-E[Y^{(z)}\mid Z=0,X]$. If the SITA assumption holds, $c(z,X)$ equals zero. Thus, the sensitivity function can describe the magnitude of the departure from SITA assumption or the impact of the unmeasured confounders. Once we specify the sensitivity function $c(z, X)$, one can predict the mean function of the counterfactual variables conditional on $X$ and then estimate the ATE without the SITA assumption. Li et al. \cite{li2011} criticized a technical difficulty in defining the sensitivity function when covariates $X$ contain multiple dimensions. Of note, in practical sensitivity analysis, if $X$ is multi-dimensional, not only the functional form but also the specific coefficients for each covariate are required to be specified. Such specifications were criticized to be unlikely to accurately reflect the relationship between the departure from SITA assumption and the potential outcomes. Li et al. \cite{li2011} proposed a refinement by defining the sensitivity function as a function of the MAR-based propensity score:  $c(z,e(X))=E[Y^{(z)}\mid Z=1,e(X)]-E[Y^{(z)}\mid Z=0,e(X)]$. The MAR-based propensity score is a one-dimension summary of observed covariates, and this refinement made the specification of the sensitivity function much simpler. \par 
	
	%Given that the refined sensitivity function $c(z,e(X))$ can link the potential outcomes in the treated and control group, the observed outcomes can be corrected by canceling the bias owing to prescence of unmeasured confounders. Then, applying the sensitivity function-corrected observed outcomes to IPW estimators can further construct the sensitivity function-corrected IPW estimator for ATE. With specifying a plausible class of sensitivity functions, the corrected IPW estimators for ATE can be obtained, thus assessing the impacts of the unmeasured confounders. \par 
	
	However, in reality, even with the simplification by Li et al.\cite{li2011}, it is not an easy task to define a plausible range of sensitivity functions. Furthermore, their method still relies on the estimation of the MAR-based propensity score. Misspecification of the parametric model for the MAR-based propensity score may result in difficulty in interpreting the results of the sensitivity analysis.
	
	\subsection{The marginal sensitivity model}\label{sec2.32}
	Tan \cite{tan2006} proposed the marginal sensitivity model, which describes a relaxation of the SITA assumption. The model assumes a single sensitivity parameter, which permits the presence of the unmeasured confounders $U$, but restricts the extent of selection bias that can be attributed to these confounders. One can specify a parameter $\lambda$, and then the following inequality is supposed to hold:
	\begin{equation*}
		1/\lambda\le\frac{e(X_i,U_i)/(1-e(X_i,U_i))}{\hat{e}(X_i)/(1-\hat{e}(X_i))}\le \lambda,
	\end{equation*}
	where $e(X_i,U_i)$ refers to the true propensity score measuring all covariates, and $\hat{e}(X_i)$ refers to the estimated MAR-based propensity score. The single parameter $\lambda$, that is, the odds ratio (OR) between true propensity score and estimated propensity score, can control degree of unconfoundedness. When $\lambda=1$, the inclusion of additional confounders has no effect on the treatment odds. This implies that the allocation of the treatment is not influenced by confounding factors. That is, the SITA assumption holds. Increasing $\lambda$ represents the allowance for stronger extent to which the SITA assumption is violated. Tan \cite{tan2006} proposed a sensitivity analysis method to assess how the estimates based on the nonparametric likelihood change under the violation of the SITA assumption.  \par 
	
	By introducing the marginal sensitivity model, the sensitivity analysis for unmeasured confounders can be applied to the IPW estimator under the MNAR. If $U$ was observed, one can estimate $\mu_1$ with
	\begin{equation}
		\left(\frac{1}{n}\sum_{i=1}^n\frac{Z_i}{e(X_i,U_i)}\right)^{-1}\frac{1}{n}\sum_{i=1}^n\frac{Z_iY_i}{e(X_i,U_i)}.\label{eq:true}
	\end{equation}
	In practice, ${U}$ is unobserved and $\hat{\mu}_1$ in equation \eqref{eq:true} actually makes no sense. However, under the marginal sensitivity model, $\lambda$ can link the unobserved true propensity score and the estimated MAR-based propensity score, so that it is possible to evaluate bounds of \eqref{eq:true} under some constraints. That is
		\begin{equation}
			\begin{split}\label{eq:22}
	\max \mbox{or} \min &\left(\frac{1}{n}\sum_{i=1}^n\frac{Z_i}{e(X_i,U_i)}\right)^{-1}\frac{1}{n}\sum_{i=1}^n\frac{Z_iY_i}{e(X_i,U_i)}\\
	\mbox{subject to} \quad &	1/\lambda^{1}\le\frac{e(X_i,U_i)/(1-e(X_i,U_i))}{\hat{e}(X_i)/(1-\hat{e}(X_i))}\le \lambda^1,
\end{split}
		\end{equation}
	where $\lambda^1$ is the pre-specified constant, which describes the upper and lower bounds of the discrepancy of the true propensity score from the estimated MAR-based propensity score for the estimation of $\mu_1$. As long as the true propensity score for all the subjects satisfies the constraint, the true $\mu_1$ should be bounded by the minimum and maximum of \eqref{eq:22} asymptotically. It is possible to have an interval for $\mu_0$ in a similar way. This method under the marginal sensitivity model was proposed firstly by Zhao.\cite{zhao2019} In this method, the sensitivity parameter $\lambda$ quantifies the extent to which the SITA assumption is violated. However, it still suffers from defining a plausible range for the sensitivity parameter and reliance on correct specification of the MAR-based propensity score model.  \par
	
	It was criticized that the interval obtained by \eqref{eq:22} may not be tight and the interval was asymptotically conservative.\cite{dorn2022} Dorn and Guo \cite{dorn2022} proposed the quantile balancing method, a refinement based on the marginal sensitivity model. Let $F(y\mid x,z)=P(Y\le y \mid X=x,Z=z)$ and the quantile function is defined by $Q_t(x,z)=\mbox{inf}\{q:F(q\mid x,z)\ge t\}$. For bounding $\mu_1$, the quantile balancing method solves the following optimization problem:
		\begin{align}
		\label{eq:qb1}
				\max \mbox{or} \min &\left(\frac{1}{n}\sum_{i=1}^n\frac{Z_i}{e(X_i,U_i)}\right)^{-1}\frac{1}{n}\sum_{i=1}^n\frac{Z_iY_i}{e(X_i,U_i)}\\
				\mbox{subject to} \quad &\sum_{i=1}^n\left(\begin{array}{cc}
					1\\\hat{Q}_\tau(X_i,1)
				\end{array}\right)\left(\frac{Z_i}{e(X_i,U_i)}-\frac{Z_i}{\hat{e}(X_i)}\right)=0 \label{eq:qb2}\\	&1/\lambda^{1}\le\frac{e(X_i,U_i)/(1-e(X_i,U_i))}{\hat{e}(X_i)/(1-\hat{e}(X_i))}\le \lambda^1, \label{eq:qb3}
		\end{align}
	where $\tau=\frac{\lambda^1}{1+\lambda^1}$ and $\hat{Q}_\tau(X_i,1)$ is estimated with some quantile regression models.\cite{dorn2022} Bounding $\mu_0$ and the ATE can be achieved in a similar way. The quantile balancing method refined Zhao's sensitivity analysis method \cite{zhao2019} by adding the quantile function to balance the treatment assignment $Z$ over the true propensity score at population level. This additional constraint based on the estimated quantile function ensured asymptotic optimality of the interval obtained by solving \eqref{eq:qb1}. Although it solves asymptotic conservativeness in Zhao's method,\cite{zhao2019} it still suffers from the misspecification of the estimated MAR-based propensity score. Moreover, the quantile function also requires specifying some parametric models or machine learning-related methods. \par

	\section{The Proposed sensitivity analysis method} \label{sec:p}
	We begin with the bound for $\mu_1$. Let $e^1(X_i,U_i)$ denote the true propensity score for subjects in the treated group. Let us consider to construct the upper bound of $\mu_1$ by solving the following optimization problem:
		 \begin{align}
			\bar{\mu}_1^+=\quad\max\quad &\frac{1}{n}\sum_{i=1}^n\frac{Z_iY_i}{e^1(X_i,U_i)} \label{eq:24}\\
			\mbox{subject to} \quad&\delta\le e^1(X_i,U_i)\le1-\delta \label{eq:25}\\
			&\sum_{i=1}^ng(X_i)\left(1-\frac{Z_i}{e^1(X_i,U_i)}\right)=0.\label{eq:26}
		\end{align}
    In a similar way, to obtain the lower bound of $\mu_1$, let us consider the following problem: 
		 \begin{align}
			\bar{\mu}_1^-=\quad\min\quad &\frac{1}{n}\sum_{i=1}^n\frac{Z_iY_i}{e^1(X_i,U_i)} \label{eq:27}\\
			\mbox{subject to} \quad&\delta\le e^1(X_i,U_i)\le1-\delta \label{eq:28}\\
			&\sum_{i=1}^ng(X_i)\left(1-\frac{Z_i}{e^1(X_i,U_i)}\right)=0.\label{eq:29}
		\end{align}
	The constraints \eqref{eq:25} and \eqref{eq:28} come from the positivity assumption (Assumption \ref{as2}), which is a fundamental assumption in causal inference. We regard $\delta$ in \eqref{eq:25} and \eqref{eq:27} as a sensitivity parameter. The constraints \eqref{eq:26} and \eqref{eq:29} come from the estimating equation for the outcome-dependent propensity score \eqref{eq:14}. As mentioned, the estimating equation \eqref{eq:14} cannot necessarily identify the true propensity score model uniquely from a parametric model. However, according to the law of large number, it holds that 
	\begin{equation}\label{eq:asy}
		\sum_{i=1}^ng(X_i)\left(1-\frac{Z_i}{e^1(X_i,U_i)}\right)\xrightarrow{p}E\left[g(X)\left(1-\frac{Z}{e^1(X,U)}\right)\right]=0.
	\end{equation}
	Then, the true propensity scores should satisfy the constraints \eqref{eq:26} and \eqref{eq:29} asymptotically, and therefore, $\mu_1$ should be included in the interval $[\mu_1^-, \mu_1^+]$ asymptotically. The bound for $\mu_1$ is the interval $[\bar{\mu}_1^+,\bar{\mu}_1^-]$, which contains the true $\mu_1$.\par
	
	Let us consider the inverse of the true propensity score, denoted by $w_i^1=(e^1(X_i,U_i))^{-1}$, as the decision variable. Then, optimization problems \eqref{eq:24} and \eqref{eq:27} become a linear programming problem with linear constraints:
		\begin{align}
				\quad\min\mbox{or}\max\quad &\frac{1}{n}\sum_{i=1}^nZ_iY_iw_i^1 \label{eq:rm1}\\
				\mbox{subject to} \quad&\frac{1}{1-\delta}\le w_i^1\le\frac{1}{\delta} \label{eq:rm11}\\
				&\sum_{i=1}^ng(X_i)(1-Z_iw_i^1)=0. \label{eq:rm12}
		\end{align}
	Compared to the quantile balancing method, which is nonlinear optimization and requires estimation of the quantile functions, our proposal can be solved time-efficiently with the interior-point method or the simplex algorithm for the linear programming and then tractable with standard software for mathematical programming. \par 
	
	The bound for $\mu_0$ can be constructed in a similar way as follows. Let $e^0(X_i,U_i)$ denote the true propensity score for subjects in the control group and similarly consider the weight $w_i^0=(1-e^0(X_i,U_i))^{-1}$ as the decision variable. Then the interval $[\bar{\mu}_0^-,\bar{\mu}_0^+]$ can be obtained by solving the following linear programming problem:
	\begin{align}
			\quad\min\mbox{or}\max\quad &\frac{1}{n}\sum_{i=1}^n(1-Z_i)Y_iw_i^0 \label{eq:rm0}\\
			\mbox{subject to} \quad&\frac{1}{1-\delta}\le w_i^0\le\frac{1}{\delta} \label{eq:rm01}\\
			&\sum_{i=1}^ng(X_i)(1-(1-Z_i)w_i^0)=0. \label{eq:rm02}
	\end{align}
	We obtain bounds for $\psi$ by $[\bar{\mu}_1^--\bar{\mu}_0^+, \bar{\mu}_1^+-\bar{\mu}_0^-]$.\par
	
	Generally, in the estimation of propensity score, the dimension of $g(X_i)$ should be equal to the number of unknown parameters in the parametric model for the propensity score. In the proposed sensitivity analysis method, one can impose more constraints by increasing the dimension of $g(X_i)$, thereby yielding a narrower bound obtained by the linear programming problems \eqref{eq:rm1} and \eqref{eq:rm0}. $g(X_i)$ can be any function of $X_i$. Suppose that there are $K$ covariates: $X^\top=(X_{i,1},X_{i,2},\dots,X_{i,K})$. Then, $g(X_i)$ can be like
		\begin{equation}
			g(X_i)=\left(\begin{array}{cccc}
				1\\X_{i,1}\\\vdots\\X_{i,K}
			\end{array}\right) \quad\mbox{or} \quad  g(X_i)=\left(\begin{array}{ccccccc}
			1\\X_{i,1}\\\vdots\\X_{i,K}\\X_{i,1}^2\\\vdots\\X_{i,K}^2
		\end{array}\right) \quad\mbox{or}\quad g(X)=\left(\begin{array}{cccccccc}
		1\\X_{i,1}\\\vdots\\X_{i,K}\\X_{i,1}^2\\\vdots\\X_{i,K}^2\\\vdots
	\end{array}\right).\label{eq:gx}
		\end{equation}
	As long as the resulting constraints give us feasible solutions to the optimization problems, the proposed method is expected to narrow the bound by simply increasing the dimension of $g(X_i)$, since greater flexibility on the choice of $g(X_i)$ is allowed.\par 
	
	The IPW estimators \eqref{eq:ipw1} and \eqref{eq:ipw0} do not satisfy the population boundedness property: the IPW estimator can be beyond the range of the outcome.\cite{tan2010bounded,robins2007comment} On the other hand, the SIPW estimators \eqref{eq:sipw1} and \eqref{eq:sipw0} satisfy it. The objective functions \eqref{eq:24} and \eqref{eq:27} have the form of the IPW estimator. If we set the first element of $g(X_i)$ to be 1, as seen in \eqref{eq:gx}, the IPW estimator agrees with the SIPW estimator. Consequently, it is sufficient to consider using a more computationally tractable, unstabilized form as the objective function and suggested to consider 1 as the first element of $g(X_i)$.   
	
	%\begin{remark}
	%	Recall the SIPW estimator of $\mu_1$:
	%	\begin{equation}
	%		\frac{\sum_{i=1}^nZ_iY_iw_i^1}{\sum_{i=1}^nZ_iw_i^1}.
	%	\end{equation}
	%	The constraints (\ref{eq:26}) and (\ref{eq:29}) ensure that if we make $g(X_i)=1$, this leads to $\sum_{i=1}^nZ_iw_i^1=n$. Consequently, it is sufficient to consider using a more computationally tractable, unstabilized form as the objective function. Accordingly, it is suggested to consider $g(X_i)=1$ as a dimension of $g(X_i)$.\par
	%\end{remark}
	Dorn and Guo \cite{dorn2022} also considered the condition \eqref{eq:asy}, but criticized that $g(X_i)$ should involve infinitely many moment conditions. Coupled with the constraint \eqref{eq:qb3}, they showed that the infinitely many constraints can be replaced with a single constraint of the quantile balancing \eqref{eq:qb2}. This simplification with the quantile balancing is realized with the OR-based constraint \eqref{eq:qb3}. In practical sensitivity analysis of observational studies, the bounds for the ATE obtained by optimizing \eqref{eq:24} and \eqref{eq:27} are generally compared with a specific threshold, such as zero, to ensure the robustness of the results. Therefore, there is no need to introduce an infinite number of constraints, as it is sufficient to increase the dimension of $g(X_i)$ to ensure the robustness of the causal inference in an observational study. In addition, the quantile function must be estimated and then be subject to assumptions in modeling and estimation, although Dorn and Guo \cite{dorn2022} tried to minimize the risk of misspecification by introducing flexible models. The authors provides several machine learning-related estimation methods, which might yield notably different bounds from each other in their simulation study.\cite{dorn2022} One advantage of the proposed method is that it does not rely on messy estimation in the quantile regression. Simply by increasing dimension of $g(X_i)$, we can try to make the bound narrower. Furthermore, it is more crucial that  our method can provide bounds without relying of the condition \eqref{eq:qb3}, $\lambda$ in which is hard to specify. \par 
	
	On the other hand, the bounds may be wide without \eqref{eq:qb3} and may not give any meaningful information. If this is the case, the constraint \eqref{eq:qb3} can be incorporated into the optimization problems \eqref{eq:rm1} and \eqref{eq:rm0} as follows:
	\begin{align}
		\quad\min\mbox{or}\max\quad &\frac{1}{n}\sum_{i=1}^nZ_iY_iw_i^1 \label{eq:or1}\\
		\mbox{subject to} \quad&\frac{1}{1-\delta}\le w_i^1\le\frac{1}{\delta} \notag %\label{eq:34}  
  \\
		&\sum_{i=1}^ng(X_i)(1-Z_iw_i^1)=0\notag\\
		&\frac{1+(\lambda^1-1)\hat{e}(X_i)}{\lambda^1 \hat{e}(X_i)}\le w_i^1\le \frac{1+\hat{e}(X_i)(1/\lambda^1-1)}{\hat{e}(X_i)(1/\lambda^1)}. \label{eq:36}
	\end{align}
	The additional constraint \eqref{eq:36} is from the marginal sensitivity model \eqref{eq:22}, in which $\hat{e}(X_i)$ refers to estimated MAR-based propensity score depending merely on measured covariates, and $\lambda^1$ refers to the OR between true propensity score and the estimated MAR-based propensity score. Note that, after introducing the third constraint \eqref{eq:36}, the optimization problem remains a linear programming problem. With $g(X_i)$ fixed, the addition of the constraint \eqref{eq:36} can further narrow the bound obtained by solving the linear programming \eqref{eq:or1}. The bound for $\mu_0$ can be obtained in a similar way:
	\begin{equation}\label{eq:or0}
		\begin{split}
			\quad\min\mbox{or}\max\quad &\frac{1}{n}\sum_{i=1}^n(1-Z_i)Y_iw_i^0 \\
			\mbox{subject to} \quad&\frac{1}{1-\delta}\le w_i^0\le\frac{1}{\delta} \\
			&\sum_{i=1}^ng(X_i)(1-(1-Z_i)w_i^0)=0\\
			&\frac{\hat{e}(X_i)}{\lambda^0(1-\hat{e}(X_i))}+1\le w_i^0\le \frac{\lambda^0\hat{e}(X_i)}{1-\hat{e}(X_i)}+1.
		\end{split}
	\end{equation}

	As done by Dorn and Guo,\cite{dorn2022} estimation error for the bounds can be accounted by using the bootstrap confidence intervals of the lower and upper bounds. One may hope to make the bounds tighter by introducing $g(X_i)$ of the higher dimension. A concern is putting more variables may lead unreliable bounds of less stability. The bootstrap samples would also be useful to evaluate how stable the resulting bounds are: if the number of the bootstrap samples of feasible solutions of the linear programming is small, the resulting bounds should be carefully interpreted. 
	
	\section{Simulation Study}\label{sec:s}
	In this section, we investigate the performance of the proposed method and compare it with Dorn and Guo's method\cite{dorn2022} based on the marginal sensitivity model and quantile balancing over several simulated datasets. The simulation settings followed Morikawa and Kim's framework,\cite{morikawa2021} allowing us to evaluate the performance of the proposed method when encountering unidentifiability issues. In our simulation, we considered to generate five covariates $\bar{X}_i^\top=(X_{i,1},X_{i,2},X_{i,3},X_{i,4},X_{i,5})$ from the normal distribution with
		\begin{equation*}
			\begin{split}
				X_{i,1}&\sim \mathcal{N}\left(0,1\right)\\
				X_{i,k+1}\mid X_{i,k}=x_{i,k}&\sim \mathcal{N}\left(\frac{-x_{i,k}}{3},1\right), \quad k=1,2,3,4.
			\end{split}
		\end{equation*}
	Here, $\{X_{i,1},X_{i,2},X_{i,3},X_{i,4}\}$ were regarded as measured covariates, while $X_{i,5}$ was regarded as an unmeasured confounder. By setting different $a_1^{[s]}$ in two scenarios $(s=1,2)$, where $a_1^{[1]}=0.0775$ and $a_1^{[2]}=0.998$, the outcome was generated as follows:
		  \begin{equation*}
			\mu^{(1)}(\bar{x}_i)=a_1^{[s]}+0.4x_{i,1} +0.4x_{i,2} +0.6x_{i,1}x_{i,2} +0.5x_{i,3}-0.7x_{i,4} +0.2x_{i,5}, 
		\end{equation*}
		\begin{equation*}
			\mu^{(0)}(\bar{x}_i)=0.0654+0.2x_{i,1}+0.1x_{i,2}+1.2x_{i,1}x_{i,2}+0.2x_{i,3}-0.3x_{i,4}+0.6x_{i,5},
		\end{equation*}
		\begin{equation*}
			Y^{(1)}_i\mid (\bar{X}_i=\bar{x}_i)\sim \mathcal{N}\left(\mu^{(1)}(\bar{x}_i),\frac{1}{4}\right),
		\end{equation*}
		\begin{equation*}
			Y^{(0)}_i\mid (\bar{X}_i=\bar{x}_i)\sim \mathcal{N}\left(\mu^{(0)}(\bar{x}_i),\frac{1}{4}\right).
		\end{equation*}
	The treatment assignment $Z_i\in\{0,1\}$ was generated by the Bernoulli distribution with
	\begin{equation*}
		P(Z_i=1\mid \bar{X}_i=\bar{x}_i,Y^{(1)}_i=y^{(1)}_i)=\frac{1}{1+\exp{(-0.904+0.5x_{i,1}+0.5x_{i,2}+0.5x_{i,3}-0.2x_{i,4}-x_{i,5}+0.3y^{(1)}_i)}}.
	\end{equation*}
	We simulated 1,000 observational studies with $n=1,000$ subjects for each scenario. For each simulated study, the bounds for the ATE were calculated by our proposed method, as well as the quantile balancing method.\cite{dorn2022} In applying the quantile balancing method, the linear quantile regression on $\{X_{i,1},X_{i,2},X_{i,3},X_{i,4}\}$ was applied. For the constraint \eqref{eq:qb3}, we estimated the MAR-based propensity score $\hat{e}(X_i)$ by the logistic regression model with $\{X_{i,1},X_{i,2},X_{i,3},X_{i,4}\}$ and applied 5-fold cross-fitting in the estimation of the quantile function. The application of the quantile balancing method was conducted by utilizing the R package provided by Dorn and Guo.\cite{dorn2022} The proposed method with \eqref{eq:rm1} and \eqref{eq:rm0} did not involve any estimation of the MAR-based propensity score. The OPTMODEL Procedure of SAS (SAS Institute Inc, Cary, North Carolina) was used for solving the linear programming problems in the proposed method. We considered four settings of the specification of $g(X)$: 
	\begin{enumerate}[D1]
		\item includes $1$, and the linear and quadratic terms for all the observed covariates; 
		\item includes D1 plus all two-variable interactions; 
		\item includes D1 plus the cubic terms for all the observed covariates and all interactions;
		\item includes D3 plus the quartic and quintic terms for all the observed covariates, $X_{i,4}^3(X_{i,3}-X_{i,2})(X_{i,1}+3X_{i,2}/2)$, and $X_{i,4}^3(X_{i,3}-X_{i,2})/(X_{i,1}-X_{i,3})(X_{i,1}+3X_{i,2}/2)$.
	\end{enumerate}\par
	
\begin{figure}
		\centering
		\subfloat[The proposed method ($\delta=0.01$) ]{\label{a1}\includegraphics[width=8cm]{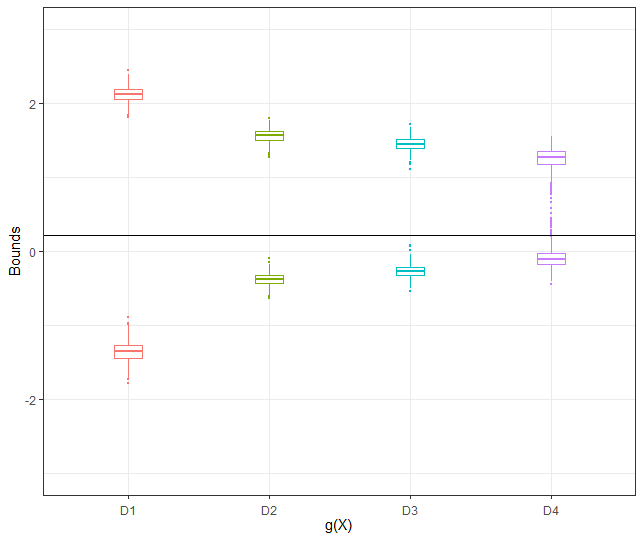}}
		\subfloat[The quantile balancing method]{\label{a2}\includegraphics[width=8cm]{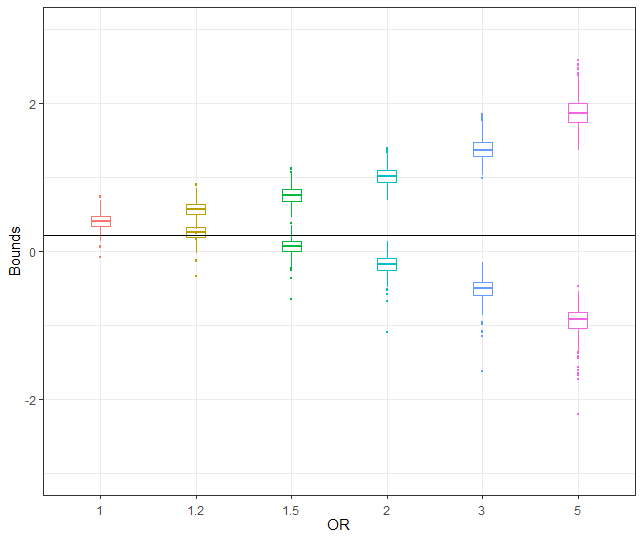}}
		\caption{Boxplots of the lower and upper bounds for ATE obtained by the proposed method (left panel) and the quantile balancing method (right panel) over 1,000 simulated datasets in Scenario 1}
		\label{fig1}
	\end{figure}

\begin{figure}
		\centering
		\subfloat[The proposed method ($\delta=0.01$)]{\label{b1}\includegraphics[width=8cm]{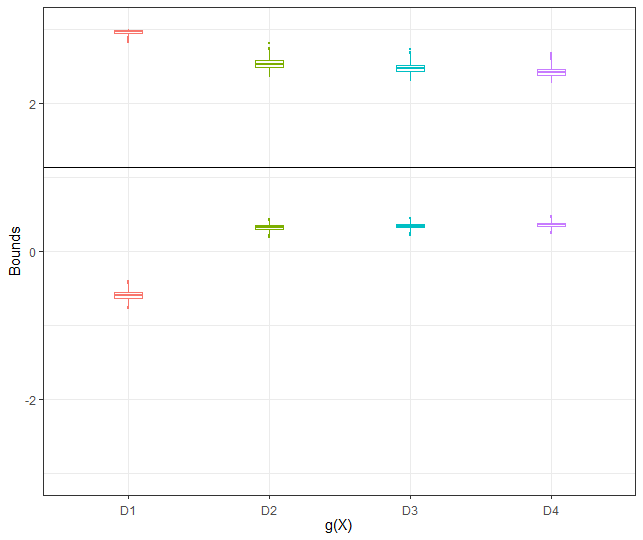}}
		\subfloat[The quantile balancing method]{\label{b2}\includegraphics[width=8cm]{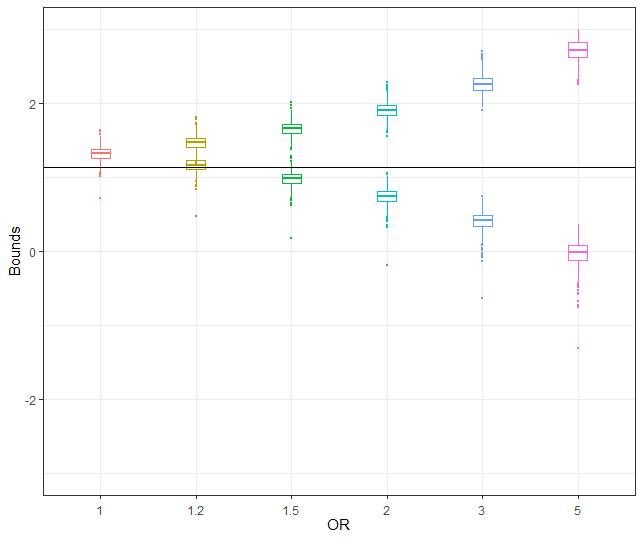}}
		\caption{Boxplots of the lower and upper bounds for ATE obtained by the proposed method (left panel) and the quantile balancing method (right panel) over 1,000 simulated datasets in Scenario 2}
		\label{fig2}
	\end{figure}

	In the proposed method, we considered to estimate the bounds for $\mu_1$ under the constraints \eqref{eq:rm11} and \eqref{eq:rm12} and for $\mu_0$ under the constraints \eqref{eq:rm01} and \eqref{eq:rm02}. Thus, the validity of the proposed method would be dependent on the choice of $\delta$. To discuss this point, we checked the distributions of the true propensity score in the simulated datasets. In the datasets under Scenario 1, with $\delta=0.1, 0.01$, and $0.001$, 18.39\%, 0.29\% and 0.01\% true propensity scores did not satisfy the conditions \eqref{eq:rm11} and \eqref{eq:rm01}, respectively. The corresponding proportions for the datasets under Scenario 2 were 2.47\%, 0.01\% and 0.00\%, respectively. Thus, with $\delta=0.1$, the constraints seemed not to hold, whereas setting $\delta=0.01$ or less were relevant. Table \ref{tab1} shows the average of the upper and the lower bounds and the coverage probability of the bounds for the ATE in the proposed method with several settings of $\delta$ and $g(X_i)$, where the coverage probability was defined as the proportion that the lower and the upper bounds covered the true ATE. The left panels of Figure \ref{fig1} and Figure \ref{fig2} show the boxplots of the lower and upper bounds for the ATE when $\delta$ was set to 0.01 in the two scenarios, respectively. We computed the averages of $Y^{(1)}$ and $Y^{(0)}$ over all simulated datasets and regarded their subtraction as the true ATE, which is depicted by a solid horizontal line in Figure \ref{fig1} and Figure \ref{fig2}. In Scenario 1, as shown in Table \ref{tab1} and the left panel of Figure \ref{fig1}, the proposed method demonstrated excellent performance in terms of the coverage probability when $\delta$ was set to 0.01 or smaller. For Scenario 2, as shown in Table \ref{tab1} and the left panel of Figure \ref{fig2}, the coverage probability of the bounds was even excellent when $\delta$ was set to 0.1. In addition, the bounds (D2, D3, and D4) in Scenario 2 effectively excluded the null. \par
    For the quantile balancing method, the right panels of Figure \ref{fig1} and Figure \ref{fig2} present the boxplots of the bound for the ATE based on different ORs in the two scenarios, respectively; the detailed estimates are summarized in Table \ref{tab2}. As the decrease of OR, the quantile balancing method gave a narrower bound with sacrificing the coverage probability. The proposed method provided feasible solutions for all the 1,000 simulated datasets with different settings of $\delta$ and $g(X_i)$, except for one setting (Scenario 2, D4, and $\delta=0.1$), in which the coverage probability was almost 1 among 952 simulated datasets with feasible solutions and the bounds for ATE were the narrowest. In this case, the proposed method gave the worst-case bounds with an average length of 1.09, which was less than the length of the bound obtained by assuming OR to be 2 in the quantile balancing method. In words, by increasing the dimension of the function alone, we can narrow down the length of the bound obtained by the proposed method to the level of quantile balancing method with OR specified as 2. The results comply with our expectations that (1) the worst-case bound obtained by our proposal can cover the true ATE without any additional assumptions; and (2) by increasing the dimension of $g(X_i)$, narrowing of our bound can be achieved. \par
	
	\section{Application}\label{sec:a}
	In this section, we apply the proposed method to a real-world data from TONE study.\cite{sasaki2009prevalence} This study aimed to evaluate the effectiveness of a designated exercise program in preventing dementia among the elderly. In this study, scores in five cognitive domains (attention, memory, visuospatial function, language, and reasoning) were used to quantify the level of cognition. We considered to estimate the effectiveness of the exercise program on the attention domain, which was regarded as a continuous variable. The confounders included age, sex, education level ($1/0$: high/low), and attention scores at the baseline.\par
	
%	\begin{table*}%
%		\centering
%		\caption{Summary of the standardized difference between the treated and the control groups with and without the inverse weighting\label{tab3}}
%		\begin{threeparttable}
%			\begin{tabular}{ccc}
%				\hline
%				& \multicolumn{2}{c}{Standardized difference } \\ \cline{2-3} 
%				Covariates                   & Unweighted                               & Weighted                              \\ \hline
%				Age                          & -0.43                                   & -0.08                                 \\
%				Gender                       & -0.02                                    & 0.04                                  \\
%				Attention scores at baseline & 0.77                                    & 0.03                                  \\
%				Education level              & 0.63                                     & 0.03                                  \\ \hline
%			\end{tabular}
%		\end{threeparttable}
%	\end{table*}
	
	In the primary analysis, a total of 935 participants were included, in which 234 were in the exercise program group and 701 in the control group. We utilized the IPW estimator to adjust imbalances in covariates between the exercise program and control groups. In the primary analysis, we used the logistic regression to the estimate MAR-based propensity scores, and the unknown parameters were estimated by the maximum likelihood method. Significantly large between-group imbalances were observed in age, attention scores at baseline, and education level before weighting inversely by the estimated MAR-based propensity scores. The between-group imbalances were effectively eliminated after weighting, indicating that IPW significantly enhanced balance across the two groups. The IPW point estimate of the ATE was 4.09 with a 95\% confidence interval of $[2.97,5.22]$. Therefore, the result of primary analysis indicated a significantly positive effect of the exercise program on the improvement of the attention level. This finding was consistent with some previous randomized controlled trials.\cite{sanders2020,lamb2018} However, a meta-analysis of observational studies\cite{demurtas2020} reported an insignificantly positive result, suggesting that the robustness of the positive effects needs further confirmation. Then, a sensitivity analysis was necessary.
	
	In the sensitivity analysis, we applied our proposed method with $\delta=0.01$ and four settings of $g(X_i)$:
	\begin{enumerate}[D1]
		\item includes $1$ and the linear term of all the covariates;
		\item includes $1$ and the linear and quadratic terms of all the covariates; 
		\item includes D2 plus the interaction between age and attention score at baseline;
		\item includes D2 plus all two-variable interactions.
	\end{enumerate}
	For the quantile balancing method, the quantile function was estimated using linear quantile regression with 5-fold cross fitting, and the MAR-based propensity scores were estimated using the logistic regression. The result by the proposed method is given in Table \ref{tab4}. In addition to the plain bounds, we calculated the confidence intervals of the upper and lower bounds with 1,000 bootstrap samples. BootLower and BootUpper refer to the lower and upper bounds of 95\% bootstrap confidence interval for the lower and upper bounds of ATE, respectively. Feasibility refers to the number of the resampled datasets, in which feasible solutions of the linear programming in the proposed method can be obtained. At first, we examined the worst-case bounds based on the proposed method \eqref{eq:rm1} without the OR-based constraints (Table \ref{tab4}). The resulting bounds with the four settings of $g(X_i)$ are presented in the row with $\lambda=/$ in Table \ref{tab4}. Even with $g(X_i)$ of a higher dimension (D4), the lower bound was less than the null value 0, indicating that the proposed methods did not eliminate concerns on unmeasured confounders. Since the length of the bounds was wide and not necessarily well interpretable, the OR-based constrains were added, and the bounds were calculated with \eqref{eq:or1} and \eqref{eq:or0}. The results with $\lambda=2, 3$, and $5$ are also shown in Table \ref{tab4}. Our proposal indicated that the worst-case bound could exclude the null if it was supposed that the OR between the true propensity score and estimated MAR-based propensity score was at most 2. For reference, we also applied the quantile balancing method. The corresponding bounds with the OR-based constrains based on the quantile balancing method \eqref{eq:qb1} are presented in Table \ref{tab5}. The bounds were similar to ours and, of note, when subjected to the same OR-based constraint, our bound was tighter than that of the quantile balancing method. We noticed that the smaller OR and greater complexity of $g(X_i)$ caused less feasible solutions of the linear programming in resampled datasets. Without the additional OR=based constraint, the results of bootstrap kept stable and feasible. However, when a small OR was assumed, occasions to have feasible solutions in resampled datasets drastically decreased. Thus, the estimating equation constraints are not necessarily compatible with the OR-based constraints, in particular when a small $\lambda$ is set. \par

	By incorporating more covariates, we tried to make the bounds tighter. We further included the baseline scores of other four cognitive domains (memory, visuospatial function, language, and reasoning) as confounders in the sensitivity analysis. The results by the proposed method without the OR-based constraint \eqref{eq:rm1} are shown in Table \ref{tab6}. The worst-case bounds for the ATE in Table \ref{tab6} achieve great tightness. In some settings (D3 and D4), the worst-case bounds even excluded the null, thus indicating the robustness of the primary analysis, without any additional OR-based constraints. Nevertheless, we noticed small numbers of boostrap samples with feasible solutions. Thus, the successful exclusion of the null with D3 and D4 would be subject to instability, and it could not eliminate concerns against unmeasured confounders completely. \par  
	
	\section{Discussion}\label{sec:d}

	Interest in drawing medical evidence from the real-world data has been rapidly growing, and the number of papers reporting results of real-world data analyses with the confounder adjustment has been substantially increasing. The propensity score analysis is now routinely applied in the analysis of observational studies. However, almost all the papers only reports the results of the propensity score matching and/or the IPW method by the propensity score and do not address the important issue of the unmeasured confounders. Since the issue of residual confounding is always left as a limitation in the analysis of observational studies, it is very important to develop sensitivity analysis methods which are easily applicable and rely on less assumptions. In this paper, we proposed a simple sensitivity analysis method based on the IPW method. To our best knowledge, all the existing sensitivity analysis methods for the IPW estimator rely on some untestable assumptions on the departure from the SITA assumption. Although they provide very useful tools to address potential impacts of unmeasured confounders, there are still concerns with potential violation of the assumption. Our method requires only minimal assumptions and can construct the bounds for the ATE, completely free from any quantification of the departure from the SITA assumption. Although it may not give sufficiently informative bounds of small width, showing the bounds based on minimal assumptions would be useful as a basis in addressing the potential impacts of the residual confounding. Our method can easily incorporate the OR-based constraints for the departure from the SITA assumption to give tighter bounds. The resulting method with the additional constraints corresponds to the quantile methods by Dorn and Gou.\cite{dorn2022} Our method can be easily applied with the linear programming, avoiding the estimation of the quantile functions, and empirically, gave a bit tighter bounds in our simulation study. We propose a strategy that analysts begin with the bounds without the OR-based constraints of minimal assumptions and then incorporates additional OR-based constraints if needed. Comparing with the elegant theory of the quantile balancing method by Dorn and Guo,\cite{dorn2022} our method is based on a very simple idea. We believe that our method is practical and our strategy would be useful in addressing the issue of residual confounding. \par 
	
	By incorporating more constraints with $g(X)$ of higher dimension, one may have tighter bounds with our proposed method. It motivates us to collect as many potential confounders as possible in conducting observational studies. On the other hand, we do not have any clear guidance on how to define $g(X)$. Putting more constraints would be desirable to make the bounds tighter. However, feasibility depends on sample size, and we cannot implement the linear programming without any feasible solution which depends on the choice of $g(X)$. Establishing a practical guidance would be an important future research topic. \par 
	
	Our idea is simple: we remove any parametric models for the propensity score, but still rely on the estimating equation for the propensity score. This simplicity would make us easily extend the idea to more complicated problems in causal inference and missing data analysis. This also warrants addressing in future research. 
	
	\section*{Acknowledgments}
	
	This research was partly supported by Grant-in-Aid for Challenging Exploratory Research (16K12403) and for Scientific Research(16H06299, 18H03208) from the Ministry of Education, Science, Sports and Technology of Japan.

    	\section*{Conflict of interest}

     The authors declare no conflict of interests.

	%\section*{Data available statement}
	%R codes together with a sample application data set are available at
	%\url{xxx}.
	
	%\section*{Supporting information}
	%Additional supporting information can be found online in the Supporting Information section at the end of this article.

\bibliography{Reference}

\newpage

	\begin{table*}%
		\centering
		\caption{Summary of the proposed method with several settings of $\delta$ and $g(X)$ over 1,000 simulated datasets in two scenarios.\label{tab1}}
		\begin{threeparttable}
		\begin{tabular}{cccccccc}
			\hline
			\multicolumn{1}{l}{} & \multicolumn{1}{l}{} & \multicolumn{3}{l}{Scenario 1 (True ATE: 0.21)} & \multicolumn{3}{l}{Scenario 2 (True ATE: 1.13)} \\ \hline
			$g(X)$ & $\delta$ & Bound$^{*}$            & Length$^{**}$ & Coverage$^{***}$ & Bound            & Length & Coverage  \\ \hline
			D1   & 0.1$^{a}$   & {[}-0.44,1.41{]} & 1.85   & 1.00         & {[}0.32,2.41{]}  & 2.09   & 1.00          \\
			& 0.01$^{b}$  & {[}-1.36,2.12{]} & 3.48   & 1.00         & {[}-0.60,3.09{]} & 3.68   & 1.00          \\
			& 0.001$^{c}$ & {[}-1.43,2.18{]} & 3.61   & 1.00         & {[}-0.93,3.18{]} & 4.11   & 1.00          \\ \hline
			D2   & 0.1   &       {[}0.10,1.08{]}           &   0.97     & 0.93     & {[}0.90,2.07{]}  & 1.17   & 1.00          \\
			& 0.01  & {[}-0.38,1.56{]} & 1.94   & 1.00         & {[}0.32,2.53{]}  & 2.21   & 1.00          \\
			& 0.001 & {[}-0.41,1.59{]} & 2.00   & 1.00         & {[}0.05,2.57{]}  & 2.53   & 1.00          \\ \hline
			D3   & 0.1   & {[}0.14,1.04{]}  & 0.91   & 0.69     & {[}0.91,2.05{]}  & 1.14   & 1.00          \\
			& 0.01  & {[}-0.33,1.50{]} & 1.83   & 1.00         & {[}0.34,2.47{]}  & 2.13   & 1.00          \\
			& 0.001 & {[}-0.35,1.53{]} & 1.88   & 1.00         & {[}0.09,2.50{]}  & 2.41   & 1.00          \\ \hline
			D4   & 0.1   & {[}0.21,0.97{]}  & 0.75   & 0.45     & {[}0.93,2.02{]}  & 1.09   & 0.99 \\
			& 0.01  & {[}-0.19,1.37{]} & 1.56   & 0.95        & {[}0.36,2.43{]}  & 2.07   & 1.00          \\
			& 0.001 & {[}-0.21,1.38{]} & 1.59   & 0.97        & {[}0.13,2.45{]}  & 2.32   & 1.00         \\ \hline
		\end{tabular}
		\begin{tablenotes}
			%\centering
			\footnotesize
            \item [*] the averages of the lower and upper bounds;
            \item [**] the difference between the averages of the lower and upper bounds;
            \item [***] the proportion of inclusion of the true ATE between the lower and upper bounds;
			\item[a] 18.39\% and 2.47\% of the true propensity scores did not satisfy the constraints \eqref{eq:rm11} or \eqref{eq:rm01}, respectively, in scenarios 1 and 2;
			\item[b] 0.29\% and 0.01\% of the true propensity scores did not satisfy the constraints \eqref{eq:rm11} or \eqref{eq:rm01}, respectively, in scenarios 1 and 2;
			\item[c] 0.01\% and 0.00\% of the true propensity scores did not satisfy the constraints \eqref{eq:rm11} or \eqref{eq:rm01}, respectively, in scenarios 1 and 2.
		\end{tablenotes}
		\end{threeparttable}
	\end{table*}

	\begin{table*}%
		\centering
		\caption{Summary of the quantile balancing method with several settings of OR over 1,000 simulated datasets in two scenarios.\label{tab2}}
        \begin{threeparttable}
		\begin{tabular}{ccccccc}
			\hline
			& \multicolumn{3}{c}{Scenario 1 (True ATE: 0.21)} & \multicolumn{3}{c}{Scenario 2 (True ATE: 1.13)} \\ \hline
			$\lambda$  & Bound$^{*}$             & Length$^{**}$  & Coverage$^{***}$  & Bound            & Length & Coverage \\ \hline
			1   & {[}0.01,0.01{]}  & /      & /        & {[}1.31,1.31{]}  & /      & /        \\
			1.2 & {[}-0.13,0.14{]} & 0.27   & 0.48     & {[}1.16,1.47{]}  & 0.31   & 0.36     \\
			1.5 & {[}-0.29,0.31{]} & 0.61   & 0.89     & {[}0.98,1.66{]}  & 0.68   & 0.96    \\
			2   & {[}-0.51,0.54{]} & 1.04   & 1.00         & {[}0.74,1.90{]}  & 1.17   & 1.00         \\
			3   & {[}-0.81,0.87{]} & 1.68   & 1.00         & {[}0.40,2.26{]}  & 1.85   & 1.00         \\
			5   & {[}-1.22,1.35{]} & 2.57   & 1.00         & {[}-0.03,2.74{]} & 2.77   & 1.00         \\ \hline
		\end{tabular}
        \begin{tablenotes}
			%\centering
			\footnotesize
            \item [*] the averages of the lower and upper bounds;
            \item [**] the difference between the averages of the lower and upper bounds;
            \item [***] the proportion of inclusion of the true ATE between the lower and upper bounds.
		\end{tablenotes}
		\end{threeparttable}
	\end{table*}

	\begin{table*}%
		\centering
		\caption{Bounds of the ATE for the attention score in TONE study by the proposed method.\label{tab4}}
        \begin{threeparttable}
		\begin{tabular}{cccccccc}
			\hline
			$g(X)$ & $\lambda$ & Lower bound & Upper bound & Length & BootLower$^{*}$ & BootUpper$^{*}$ & Feasibility$^{**}$ \\\hline
			D1 & /$^{***}$  & -10.57 & 18.30 & 28.87 & -11.50 & 19.27   & 1000 \\
			& 2 & 0.57   & 6.61  & 6.04  & -0.36  & 7.59    & 987  \\
			& 3 & -1.18  & 8.59  & 9.77  & -2.39  & 9.48    & 1000 \\
			& 5 & -3.36  & 10.77 & 14.14 & -4.69  & 11.72   & 1000 \\ \hline
			D2    & /  & -9.81       & 17.75       & 27.56  & -10.83    & \multicolumn{1}{c}{18.39}     & 1000     \\
			& 2 & 1.66   & 5.50  & 3.84  & 0.05   & 6.93    & 386  \\
			& 3  & -0.92       & 8.18        & 9.10   & -1.82     & \multicolumn{1}{c}{8.92}      & 892      \\
			& 5 & -3.19  & 10.38 & 13.57 & -4.12  & 11.12   & 997  \\ \hline
			D3 & / & -9.70  & 17.23 & 26.93 & -10.46 & 17.49   & 1000 \\
			& 2 & /      & /     & /     & 0.42   & 6.73    & 66   \\
			& 3 & -0.40  & 7.45  & 7.85  & -1.41  & 8.43    & 593  \\
			& 5 & -3.08  & 9.91  & 13.00 & -3.67  & 10.59   & 957  \\ \hline
			D4 & / & -8.74  & 16.54 & 25.28 & -8.26  & 16.17   & 991  \\
			& 2 & /      & /     & /     & 0.85   & 6.39    & 55   \\
			& 3 & -0.25  & 7.16  & 7.41  & -0.86  & 7.85    & 540  \\
			& 5 & -2.66  & 9.48  & 12.15 & -2.92  & 9.93 & 920  \\ \hline
		\end{tabular}
        \begin{tablenotes}
			%\centering
			\footnotesize
            \item [*] the lower and upper bounds of 95\% bootstrap confidence interval for the lower and upper bounds of ATE, respectively;
            \item [**] the number of the resampled datasets in the proposed method, in which feasible solution of the linear programming can be obtained;
            \item [***] the rows with $/$ in the $\lambda$ column show the results without OR-based constraint.
		\end{tablenotes}
		\end{threeparttable}
	\end{table*}
	
	\begin{table*}%
		\centering
		\caption{Bounds of the ATE for the attention score in TONE study by the quantile balancing method.	\label{tab5}}
        \begin{threeparttable}
		\begin{tabular}{llllll}
			\hline
			$\lambda$  & Lower bound & Upper bound & Length & BootLower$^{*}$ & BootUpper$^{*}$ \\\hline
			1   & 4.09        & 4.09        & /      & 3.04      & 5.30      \\
			1.2  & 3.29        & 4.92        & 1.63   & 2.21      & 6.14      \\
			1.5 & 2.32        & 5.97        & 3.64   & 1.20      & 7.21      \\
			2   & 1.13        & 7.29        & 6.17   & -0.01     & 8.59      \\
			3   & -0.61       & 9.15        & 9.77   & -1.85     & 10.58     \\
			5   & -2.80       & 11.56       & 14.36  & -4.25     & 13.27     \\\hline
		\end{tabular}
         \begin{tablenotes}
			%\centering
			\footnotesize
            \item [*] the lower and upper bounds of 95\% bootstrap confidence interval for the lower and upper bounds of ATE, respectively.
		\end{tablenotes}
		\end{threeparttable}
	\end{table*}

	\begin{table*}%
		\centering
		\caption{Bounds of the ATE for the attention score in TONE study by the proposed method with $g(X)$ of additional four domains. \label{tab6}}
        \begin{threeparttable}
		\begin{tabular}{lllllll}
			\hline
			$g(X)$ & Lower bound & Upper bound & Length & BootLower$^{*}$ & BootUpper$^{*}$ & Feasibility$^{**}$ \\ \hline
			D1    & -6.75       & 14.93       & 21.68  & -8.84     & 15.84     & 1000        \\
			D2    & -2.08       & 11.14       & 13.22  & -4.29     & 12.44     & 657         \\
			D3    & 0.31        & 8.32        & 8.01   & -3.14     & 11.33     & 323         \\
			D4    & 0.70        & 7.82        & 7.12   & -2.45     & 10.51     & 220         \\ \hline
		\end{tabular}
        \begin{tablenotes}
			%\centering
			\footnotesize
            \item [*] the lower and upper bounds of 95\% bootstrap confidence interval for the lower and upper bounds of ATE, respectively;
            \item [**] the number of the resampled datasets in the proposed method, in which feasible solution of the linear programming can be obtained.
		\end{tablenotes}
		\end{threeparttable}
	\end{table*}

\newpage

\end{document}